\documentclass[12pt]{iopart}
\usepackage{graphicx}

\begin{document}

\title[]{Roles of Non-switchable Domains and Internal Bias in Electrocaloric and Pyroelectric effects}

\author{Jun Usami$^{*}$, Yuki Okamoto, Hisashi Inoue, Takeshi Kobayashi, and Hiroyuki Yamada}

\address{National Institute of Advanced Industrial Science and Technology (AIST), 1-1-1 Higashi, Tsukuba, Ibaraki 305-8565, Japan}
\ead{j-usami@aist.go.jp}
\vspace{10pt}
\begin{indented}
\item[]June 2025
\end{indented}

\begin{abstract}

Solid-state cooling and energy harvesting via pyroelectric effect (PEE) and electrocaloric effect (ECE) in ferroelectric thin films could be enhanced beyond their intrinsic ferroelectric response by exploiting the recently observed direction-dependent enhancement of the PEE; however, its microscopic origin remains unknown.
Herein, we report direct hysteresis measurements of pyrocurrent ($I_{\rm p}$) and ECE-induced temperature change versus bias voltage in 1-$\mu$m-thick Pb(Zr$_{0.65}$Ti$_{0.35}$)O$_3$ capacitors.
Both hysteresis loops exhibit pronounced asymmetries along the voltage and response axes.
By superimposing direct current voltage offsets, we isolate a residual $I_{\rm p}$-axis shift, revealing a contribution of non-switchable ferroelectric polarization.
This non-switchable polarization can be converted into switchable polarization via poling with bipolar triangular pulses, confirming the governing role of defect-induced domain pinning.
After 100 pulses, time-dependent aging was observed for pyroelectric and electrocaloric responses, with the switchable contribution markedly decaying and the non-switchable component remaining nearly constant, indicating partial repinning.
The change in voltage-axis shift agrees well with the ratio of non-switchable to switchable polarization, demonstrating that voltage shift also arises from pinned domains.
These insights clarify the critical role of non-switchable polarization in the PEE and ECE performance, suggesting strategies to optimize the directional response in ferroelectric devices through controlled poling and defect engineering.

\end{abstract}

%
%
%
%
%

\section{Introduction}

Ferroelectric materials exhibiting pyroelectric effect (PEE) and electrocaloric effect (ECE) are promising for next-generation devices, such as sensors, energy harvesters, and solid-state cooling systems~\cite{Whatmore1986,Zhang2021,Greco2020,Torello2022}.
These unique features can be enhanced by exploiting epitaxial-strain-driven structural transitions or extrinsic effects~\cite{Karthik2011,Pandya2019a,Zhang2024}, internal bias fields~\cite{Qian2015}, and polymorphic phase transitions via compositional engineering~\cite{Feng2024a}.

Recently, we demonstrated that the PEE in ferroelectric thin films can be asymmetrically enhanced depending on the direction of direct current (DC) poling~\cite{Usami2025}.
This result implies that in addition to substrate-induced asymmetry, the PEE and ECE performance could be boosted along a preferred direction by tailoring intrinsic or fabrication-process-induced anisotropy.
Although the observed asymmetry has been attributed to non-switchable polarization, the contribution of non-switchable polarization cannot be detected using conventional polarization--voltage ($P$--$V$) hysteresis loops, where $P$ is obtained by integrating displacement current.
Therefore, the nature and origin of non-switchable polarization remain unclear.

In our previous study, we demonstrated that in pyrocurrent ($I_{\rm p}$) --$V$ hysteresis loops, non-switchable polarization correlates with the voltage-axis shift ($V_{\rm{sh}}$), suggesting a relationship between both features~\cite{Usami2025}.
Voltage shifts in $P$--$V$ hysteresis curves are widely understood as imprint effects.
To explain the origin of $V_{\rm{sh}}$, two major mechanisms have been proposed: the interfacial screening model, which directly generates an internal bias via depolarization-field compensation at the ferroelectric/electrode boundary~\cite{Grossmann2002a}, and the defect-dipole model~\cite{Lambeck1986, Warren1995}, which either produces a direct bias through the dielectric constant and polarization mismatch between embedded defect dipoles and the ferroelectric matrix~\cite{Grossmann2002a} or indirectly induces $V_{\mathrm{sh}}$ by contributing to net polarization~\cite{Pike1995,Warren1996b} or by stabilizing unidirectional polarization domain states and inhibiting switching~\cite{Robels1993}.
Importantly, a pure $V_{\rm{sh}}$ can itself yield features that mimic non-switchable polarization in $I_{\rm p}$--$V$ hysteresis curves; therefore, determining whether the observed asymmetries arise from interface effects or from either direct or indirect defect dipoles is challenging.

In this study, we report the investigation of the direct PEE and ECE on ferroelectric thin films via hysteresis loop measurements, which allowed quantifying the asymmetry along the polarization and voltage axes.
By compensating the voltage shifts, we demonstrated that polarization-axis asymmetry appears in association with non-switchable polarization.
Moreover, poling with a bipolar voltage pulse revealed that the non-switchable component originates from domain pinning, which increases the required switching field and consequently enhances the voltage shift.
These findings were observed in PEE and ECE responses, suggesting that the PEE and ECE performance can be enhanced by controlling non-switchable polarization.

\section{Experimental methods} \label{experimental}

\begin{figure}
\centering
\includegraphics[width=\columnwidth]{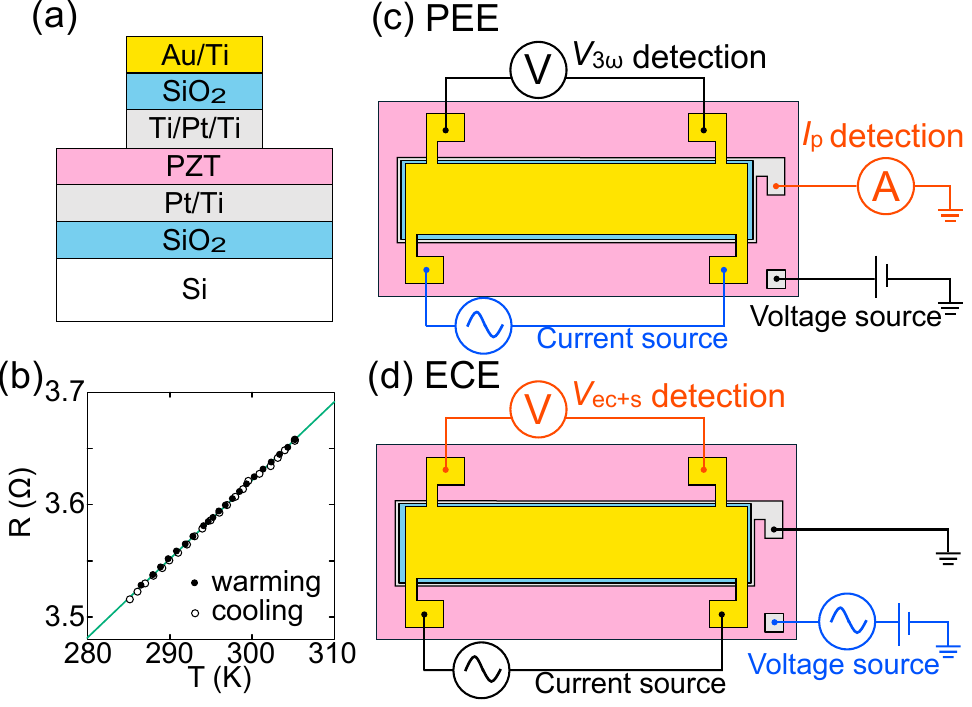}
\caption{\label{fig-setup}  (a) Cross-sectional view of the device structure. (b) Temperature dependence of the resistance of the Au/Ti heater and thermometer along with the fitted line.  (c),(d) Top views of the device structure showing the measurement setups for the PEE ($I_{\rm p}$) and ECE ($V_{\rm ec+s}$), respectively.}
\end{figure}

All experiments were conducted using ferroelectric capacitors composed of 1-µm-thick Pb(Zr$_{0.65}$Ti$_{0.35}$)O$_3$ (PZT) thin films, fabricated via radio-frequency magnetron sputtering and supplied by the Semiconductor Device Process Development Support Center~\cite{Kanda2021}.
To directly evaluate the PEE and ECE in thin films, we based our experimental approach on previous studies~\cite{Pandya2017, Usami2025} and fabricated a device structure shown in Fig.~\ref{fig-setup}(a).
The top and bottom capacitor electrodes consisted of Pt with an underlying Ti adhesion layer.
On top of the PZT capacitor structure, a SiO$_2$ insulating layer was deposited, over which an Au/Ti bilayer was patterned.
Owing to its temperature-dependent resistance, the Au/Ti bilayer served as a heater and a temperature sensor.
Figure~\ref{fig-setup}(b) displays a plot of the heater resistance versus temperature during heating and cooling, demonstrating excellent reversibility.
The temperature coefficient of resistance ($dR/dT$) was determined to be $7.00(5) \times 10^{-3}~\Omega/\mathrm{K}$.

The pyroelectric characterization was performed by applying an alternating current (AC) ($I_0=0.1$\,A, 110\,Hz) to the Au/Ti layer, generating periodic temperature oscillations in the PZT via Joule heating (Fig.~\ref{fig-setup}(c)).
The temperature oscillation amplitude $(\Delta T_{2\omega}$) was determined using the 3\(\omega\) method, which measures the third‐harmonic voltage drop (\(V_{3\omega}\)) arising from the resistance modulation at \(2\omega\) and the drive current at \(\omega\), as follows:
\begin{equation}
\Delta T_{2\omega} = \frac{2V_{3\omega}}{I_0\,(dR/dT)}.
\end{equation}
At our modulation frequency, the thermal diffusion length (\(\sim 20\)\,\textmu m) far exceeded the PZT capacitor thickness; therefore the film temperature can be regarded as uniform.
Under these conditions, \(\Delta T_{2\omega} = 0.557(6)\)\,K.
The induced pyroelectric current ($I_{\rm p}$) was detected between the Pt electrodes using a lock-in amplifier referenced to the Joule heating frequency (220\,Hz).
A DC bias voltage across the PZT film was swept to obtain $I_{\rm p}$--$V$ hysteresis loops, which enabled us to probe the field-dependent pyroelectric response.
The pyroelectric coefficient ($\Pi$) was calculated as
\begin{equation}
\Pi = \frac {I_{\rm p}}{2 \omega A \Delta T_{2\omega}},
\end{equation}
where $A$ is the heated area of the capacitor.

As shown in Fig.~\ref{fig-setup}(d), for the ECE measurements, a sinusoidal AC voltage (2\,V, $f_{\rm ec} = 18.3334$\,kHz) was superimposed on a DC bias across the PZT film.
This modulated the polarization state, generating periodic temperature oscillations via the ECE, leading to resistance oscillation of the Au/Ti heater layer at $f_{\rm ec}$.
Simultaneously, an AC current ($I_0=0.1$\,A, $f_{\rm s} = 132$\,Hz) was applied to the Au/Ti layer, and the resulting resistance change was measured as the voltage modulation at the mixed frequency of $f_{\rm ec} + f_{\rm s}$.
This approach allowed isolating the ECE signal from the direct capacitive coupling at $f_{\rm ec}$.
The mixed-frequency signal ($V_{\rm ec+s}$) was detected using a lock-in amplifier and converted to the temperature modulation amplitude ($\Delta T_{\rm sens}$) via
\begin{equation}
\Delta T_{\rm sens} = \frac{2V_{\rm ec+s}}{I_0\,(dR/dT)}.
\end{equation}
The electrocaloric coefficient was estimated using the one-dimensional heat-transport model~\cite{Pandya2017} together with the measured $\Delta T_{\rm sens}$.

Prior to each measurement, the sample was poled to stabilize its initial polarization state.
Poling was conducted by applying a 30\,V DC bias in the downward poling direction at a high temperature of 400\,K for 600 seconds, followed by slow cooling to room temperature (300\,K) over 400 seconds while maintaining the bias.
All subsequent measurements and operations were conducted at 300\,K.
The hysteresis loops were measured by sweeping the voltage  in the range of $\pm 10$\,V, which is well below the poling voltage, to minimize measurement-induced perturbation while still obtaining sufficiently clear hysteresis curves.

To evaluate the contributions from switchable and non-switchable polarization components, the responses $Y(\pm10\,\mathrm{V})$ ($Y = I_{\rm p}, \Delta T_{\rm sens}$) were analyzed.
The total response was assumed to follow the relationship $Y(\pm10\,\mathrm{V}) = \pm Y_{\rm s} + Y_{\rm ns}$, where $Y_{\rm s}$ is the switchable contribution and $Y_{\rm ns}$ is the non-switchable contribution.
$Y_{\rm s}$ and $Y_{\rm ns}$ were calculated as the half-difference and half-sum of $Y(+10\,\mathrm{V})$ and $Y(-10\,\mathrm{V})$, respectively.

\section{Results and discussion} \label{results}

\subsection{Contributions of internal bias effect and non-switchable polarization}

\begin{figure}
\centering
\includegraphics[width=0.9\columnwidth]{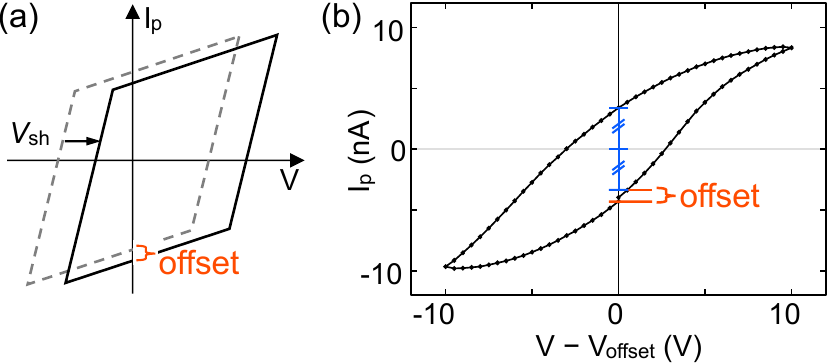}
\caption{\label{fig-Voffset}  (a) Illustration of an $I_{\rm p}$--$V$ hysteresis curve showing that a voltage-axis shift ($V_{\rm sh}$) due to internal bias alone produces an apparent offset along the $I_{\rm p}$-axis, even without any non-switchable polarization component.
(b) $I_{\rm p}$ versus $V-V_{\rm offset}$ hysteresis curve measured with $V_{\rm offset}=1.8$\,V, showing an offset in the $I_{\rm p}$-direction despite the symmetric coercive voltages.}
\end{figure}

To isolate the non-switchable polarization in the $I_{\mathrm{p}}$--$V$ loop, we superimposed a DC offset voltage ($V_{\rm offset}$) during the measurement to cancel the internal-bias-induced shift.
As shown in Fig.~\ref{fig-Voffset}(a), even in the absence of any non-switchable polarization, an internal bias alone shifts the $I_{\mathrm{p}}$--$V$ hysteresis loop along the $V$-axis, resulting in an apparent offset along the $I_{\rm p}$-axis as well.
Figure~\ref{fig-Voffset}(b) shows the $I_{\mathrm{p}}$--$V$ hysteresis curve measured under $V_{\rm offset}=1.8$\,V. 
Here, a positive bias corresponds to an upward electric field along the film normal.
Although the coercive voltages become symmetric with respect to the $V$-axis, an asymmetry along the $I_{\rm p}$-axis is still pronounced.
This residual $I_{\rm p}$-axis asymmetry confirms the presence of a non-switchable polarization component in the PZT film.

\subsection{Origin of non-switchable polarization}

Non-switchable polarization can arise from two distinct mechanisms: fixed dipoles, namely, electric polarization with a non-ferroelectric origin, and domain pinning, where the polarization switching is inhibited by defects or trapped charges.
To discriminate between these mechanisms, we varied the number of applied bipolar pulses (30\,V, 1\,kHz) after poling because bipolar pulses are known to redistribute charges and defects such as oxygen vacancies, and eliminate defect-dipoles~\cite{Glaum2012,Chen2021}.

\begin{figure}
\centering
\includegraphics[width=\columnwidth]{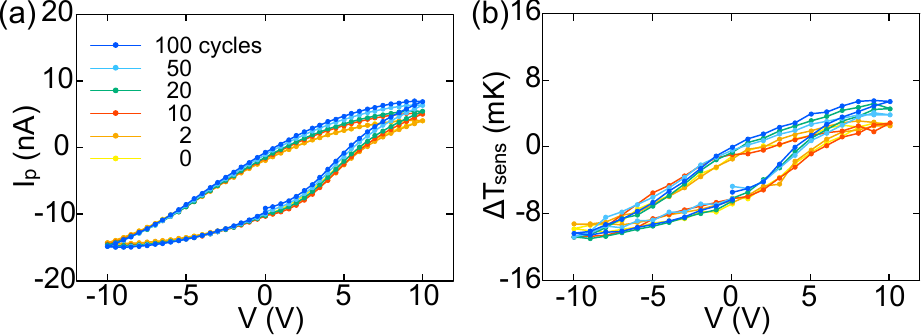}
\caption{\label{fig-bipolar}  Hysteresis curves of (a) the PEE ($I_{\rm p}$--$V$) and (b) the ECE ($\Delta T_{\rm sens}$--$V$) measured after repeated applications of  bipolar pulse.}
\end{figure}

Figures~\ref{fig-bipolar}(a) and (b) show the $I_{\rm p}$--$V$ and $\Delta T_{\rm sens}$--$V$ curves, respectively.
An asymmetric hysteresis loop was observed not only in the $I_{\rm p}$--$V$ curve but also in the $\Delta T_{\rm sens}$--$V$ curve.
In both measurements, the amplitude of the downward response was larger than that of the upward one, reflecting the initial downward poling.
The corresponding pyroelectric and electrocaloric coefficients at $-10$\,V were determined to be $-159(5)$ and $-130(10)$\,$\mu$Cm$^{-2}$K$^{-1}$, respectively.
The difference between these coefficients can be attributed to additional contributions from piezoelectric and elastocaloric effects~\cite{Tong2014a}.
With increasing the number of bipolar pulses, the downward components of the polarization response ($I_{\rm p}$ and $\Delta T_{\rm sens}$) remained almost unchanged, whereas the upward polarization responses gradually increased.
It should be noted that unlike the anomalous wake-up effect observed in HfO$_2$ films~\cite{Pandya2018b}, where the PEE coefficient increased but that of the ECE did not, our PZT films exhibited simultaneous enhancements of both PEE and ECE responses in accordance with Maxwell’s relation.

\begin{figure*}
\centering
\includegraphics[width=1\columnwidth]{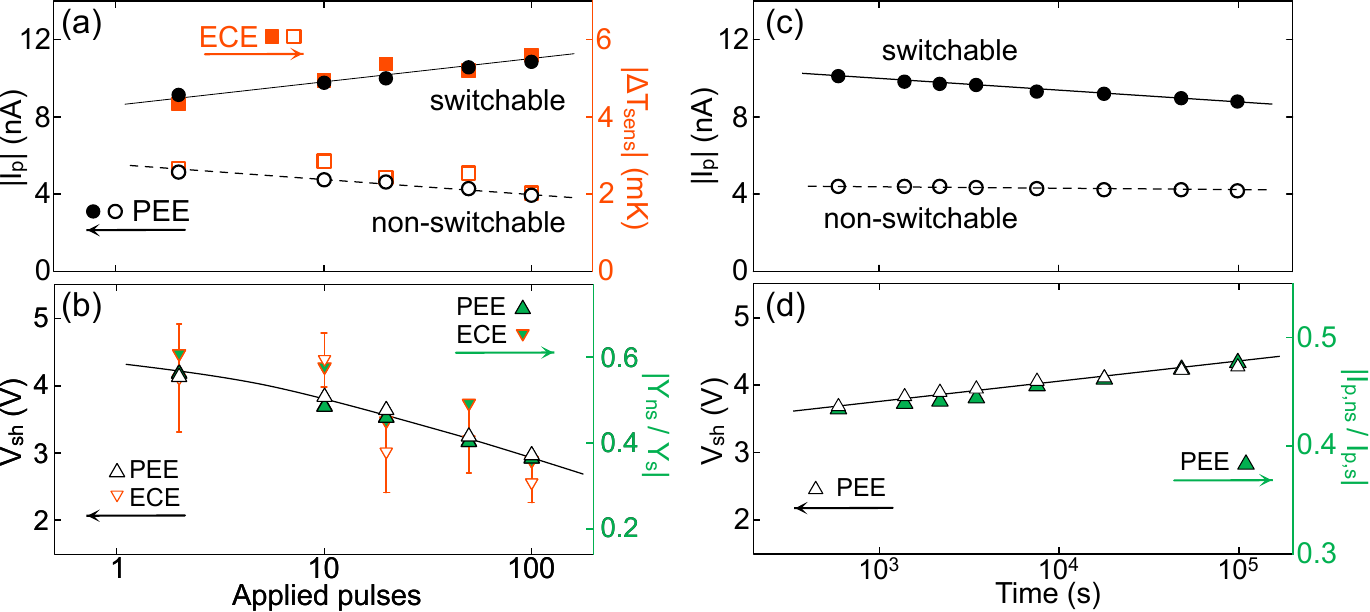}
\caption{\label{fig-dependence}  (a) Bipolar-pulse-number dependence of the absolute values of $|I_{\rm p,s}|$, $|I_{\rm p,ns}|$, $|\Delta T_{\rm sens,s}|$, and $|\Delta T_{\rm sens,ns}|$.
(b) Bipolar-pulse-number dependence of $V_{\rm sh}$ in the PEE and ECE, as compared with $|I_{\rm p,ns} / I_{\rm p,s}|$, $|\Delta T_{\rm sens,ns} / \Delta T_{\rm sens,s}|$. (c) Time evolution of $|I_{\rm p,s}|$ and $|I_{\rm p,ns}|$. (d) Time evolution of $V_{\rm sh}$ in PEE and $|I_{\rm p,ns} / I_{\rm p,s}|$.}
\end{figure*}

Figure~\ref{fig-dependence}(a) summarizes the bipolar-pulse-number dependence of the absolute values of the extracted switchable ($|I_{\rm p,s}|$, $|\Delta T_{\rm sens,s}|$) and non-switchable ($|I_{\rm p,ns}|$, $|\Delta T_{\rm sens,ns}|$) components.
As the pulse number increased, the switchable components increased, whereas the non-switchable components decreased.
For a non-switchable contribution solely due to fixed dipoles, the non-switchable component would be expected to decrease without any significant change in the switchable component.
However, the simultaneous and comparable increase in $| I_{\rm p,s}|, |\Delta T_{\rm sens,s}|$ and the decrease in $| I_{\rm p,ns}|, |\Delta T_{\rm sens,ns}|$ indicate that domain pinning is the primary origin of the non-switchable contribution.
Thus, bipolar pulses depin the ferroelectric domains, transforming non-switchable polarization into switchable polarization.

\subsection{Origin of the voltage shift in hysteresis loops}

Our previous study on pyroelectricity showed that the non-switchable polarization and the shift voltage exhibited similar poling-time dependencies~\cite{Usami2025}, although the underlying mechanisms remain unclear.
Figure~\ref{fig-dependence}(b) shows the bipolar-pulse-number dependences of $V_{\rm sh}$ and the ratio of the non-switchable and switchable components ($|I_{\rm p,ns} / I_{\rm p,s}|$, $|\Delta T_{\rm sens,ns} / \Delta T_{\rm sens,s}|$).
A simultaneous decrease in the voltage shift and the non-switchable polarization was observed, which is consistent with our previous study.

To further investigate this relationship, we monitored the time evolution of the $I_{\rm p}$--$V$ hysteresis curves after 100 bipolar pulses (30\,V, 1\,kHz).
The time evolutions of $| I_{\rm p,s}|$ and $| I_{\rm p,ns}|$ are shown in Fig.~\ref{fig-dependence}(c).
With increasing aging time, $| I_{\rm p,s}|$ gradually decreased, whereas $| I_{\rm p,ns}|$ hardly changed.
This behavior cannot be explained either by the apparent $I_{\rm p}$-axis shift induced by the $V$-axis shift or by fixed dipoles.
These results suggest an enhanced domain pinning, causing partial fixing of polarization in both upward and downward directions.

Figure~\ref{fig-dependence}(d) shows the time evolution of $V_{\rm sh}$ and $|I_{\rm p,ns} / I_{\rm p,s}|$.
Although $|I_{\rm p,ns}|$ slightly decreased over time, $V_{\mathrm{sh}}$ increased, contrary to the tendency observed in the measurements with bipolar pulse.
Conversely, $V_{\mathrm{sh}}$ showed good correlation with $|I_{\mathrm{p,ns}}/I_{\mathrm{p,s}}|$ and $|\Delta T_{\rm sens,ns} / \Delta T_{\rm sens,s}|$, indicating that the voltage shift reflects the asymmetric polarization caused by domain pinning.



\section{Conclusions} \label{conclusion}

We investigated the contributions of switchable and non-switchable polarization to the PEE and ECE responses of PZT thin-film capacitors by analyzing the $I_{\rm p}$--$V$ and $\Delta T_{\rm sens}$--$V$ hysteresis loops.
Asymmetric hysteresis shapes, which could not be attributed to internal bias alone, revealed the presence of non-switchable polarization stabilized by defect-induced domain pinning.
The application of bipolar triangular voltage pulses promoted the depinning of these domains, converting non-switchable components into switchable ones.
In contrast, time-dependent measurements showed a repinning process: the switchable polarization component decayed substantially, whereas the non-switchable component hardly changed.

The correlation between the voltage shift and the non-switchable-to-switchable polarization ratio highlights the need to distinguish these components.
Despite their different origins, both internal bias and non-switchable polarization can shift the hysteresis loops along the voltage and polarization axes, requiring careful interpretation of imprint phenomena.
Furthermore, the similar pulse-number dependences of the PEE and ECE signals confirm that PEE measurements are a reliable probe for polarization dynamics.
This study demonstrates the critical influence of non-switchable polarization on the PEE and ECE performance and offer design guidelines for optimizing ferroelectric materials and for device fabrication,  as has been previously proposed for the piezoelectric properties~\cite{Lin2024}.

\section*{Acknowledgement}
The authors gratefully acknowledge the Semiconductor Device Process Development Support Center for supplying the Pb(Zr$_{0.65}$Ti$_{0.35}$)O$_3$ thin‐film used in this study.
This work was supported by JSPS KAKENHI (Grant No. JP23K13668), JST ACT-X (Grant No. JPMJAX23K3), the Shimadzu Science Foundation, and the Suzuki Foundation.
HI acknowledges the support from JSPS KAKENHI (Grant No. JP22K14607).

\section*{References}
\bibliographystyle{unsrt}
\bibliography{All}

\begin{thebibliography}{10}

\bibitem{Whatmore1986}
R.~W. Whatmore.
\newblock {Pyroelectric devices and materials}.
\newblock {\em Reports on Progress in Physics}, 49(12):1335--1386, dec 1986.

\bibitem{Zhang2021}
Ding Zhang, Heting Wu, Chris~R. Bowen, and Ya~Yang.
\newblock {Recent Advances in Pyroelectric Materials and Applications}.
\newblock {\em Small}, 17(51):2103960, dec 2021.

\bibitem{Greco2020}
Adriana Greco and Claudia Masselli.
\newblock {Electrocaloric Cooling: A Review of the Thermodynamic Cycles,
  Materials, Models, and Devices}.
\newblock {\em Magnetochemistry}, 6(4):67, nov 2020.

\bibitem{Torello2022}
Alvar Torell{\'{o}} and Emmanuel Defay.
\newblock {Electrocaloric Coolers: A Review}.
\newblock {\em Advanced Electronic Materials}, 8(6):2101031, jun 2022.

\bibitem{Karthik2011}
J.~Karthik and L.~W. Martin.
\newblock {Pyroelectric properties of polydomain epitaxial Pb(Zr$_{1-x}$,Ti$_x$)O$_3$
  thin films}.
\newblock {\em Physical Review B - Condensed Matter and Materials Physics},
  84(2):024102, jul 2011.

\bibitem{Pandya2019a}
Shishir Pandya, Gabriel~A. Velarde, Ran Gao, Arnoud~S. Everhardt, Joshua~D.
  Wilbur, Ruijuan Xu, Josh~T. Maher, Joshua~C. Agar, Chris Dames, and Lane~W.
  Martin.
\newblock {Understanding the Role of Ferroelastic Domains on the Pyroelectric
  and Electrocaloric Effects in Ferroelectric Thin Films}.
\newblock {\em Advanced Materials}, 31(5):1803312, feb 2019.

\bibitem{Zhang2024}
S.~Zhang, J.~Deliyore-Ram{\'{i}}rez, S.~Deng, B.~Nair, D.~Pesquera, Q~Jing, M~E
  Vickers, S~Crossley, M~Ghidini, G.~G. Guzm{\'{a}}n-Verri, X.~Moya, and N.~D.
  Mathur.
\newblock {Highly reversible extrinsic electrocaloric effects over a wide
  temperature range in epitaxially strained SrTiO$_3$ films}.
\newblock {\em Nature Materials}, 23:639--647, mar 2024.

\bibitem{Qian2015}
Xiaoshi Qian, Hui‐Jian Ye, Tiannan Yang, Wen‐Zhu Shao, Liang Zhen, Eugene
  Furman, Long‐Qing Chen, and Qiming Zhang.
\newblock {Internal Biasing in Relaxor Ferroelectric Polymer to Enhance the
  Electrocaloric Effect}.
\newblock {\em Advanced Functional Materials}, 25(32):5134--5139, aug 2015.

\bibitem{Feng2024a}
Xiaoxu Feng,Ye Zhao, Yanyu Wang, Yujie Xie, Pei Han, Yong Li, and Xihong Hao
\newblock {Enhanced electrocaloric effect in KNN-based ceramic via polymorphic phase transition}.
\newblock {\em Ceramics International}, 50(1):1788--1794, 2024.

\bibitem{Usami2025}
J.~Usami, Y.~Okamoto, H.~Inoue, N.~Makimoto, T.~Kobayashi, and H.~Yamada.
\newblock {Directional enhancement of pyroelectricity in Pb(Zr$_{0.52}$Ti$_{0.48}$)O$_3$
  films by DC poling}.
\newblock {\em Applied Physics Letters}, 126(9):092902, 2025.

\bibitem{Grossmann2002a}
M.~Grossmann, O.~Lohse, D.~Bolten, U.~Boettger, T.~Schneller, and R.~Waser.
\newblock {The interface screening model as origin of imprint in PbZr$_x$Ti$_{1-x}$O$_3$
  thin films. I. Dopant, illumination, and bias dependence}.
\newblock {\em Journal of Applied Physics}, 92(5):2680--2687, sep 2002.

\bibitem{Lambeck1986}
P.V. Lambeck and G.H. Jonker.
\newblock {The nature of domain stabilization in ferroelectric perovskites}.
\newblock {\em Journal of Physics and Chemistry of Solids}, 47(5):453--461, jan
  1986.

\bibitem{Warren1995}
W.~L. Warren, D.~Dimos, G.~E. Pike, K.~Vanheusden, and R.~Ramesh.
\newblock {Alignment of defect dipoles in polycrystalline ferroelectrics}.
\newblock {\em Applied Physics Letters}, 67(12):1689--1691, sep 1995.

\bibitem{Pike1995}
G.~E. Pike, W.~L. Warren, D.~Dimos, B.~A. Tuttle, R.~Ramesh, J.~Lee, V.~G.
  Keramidas, and J.~T. Evans.
\newblock {Voltage offsets in (Pb,La)(Zr,Ti)O$_3$ thin films}.
\newblock {\em Applied Physics Letters}, 66(4):484--486, jan 1995.

\bibitem{Warren1996b}
W.~L. Warren, G.~E. Pike, D.~Dimos, K.~Vanheusden, H.N. Al-Shareef, B.~A.
  Tuttle, R~Ramesh, and J~T Evans.
\newblock {Voltage Shifts and Defect-Dipoles in Ferroelectric Capacitors}.
\newblock {\em MRS Proceedings}, 433:257, feb 1996.

\bibitem{Robels1993}
U.~Robels and G.~Arlt.
\newblock {Domain wall clamping in ferroelectrics by orientation of defects}.
\newblock {\em Journal of Applied Physics}, 73(7):3454--3460, apr 1993.

\bibitem{Kanda2021}
Kensuke Kanda, Tsukasa Koyama, Takeshi Yoshimura, Shinichi Murakami, and
  Kazusuke Maenaka.
\newblock {Characteristics of Sputtered Lead Zirconate Titanate Thin Films With
  Different Layer Configurations and Large Thickness}.
\newblock {\em IEEE Transactions on Ultrasonics, Ferroelectrics, and Frequency
  Control}, 68(5):1988--1993, may 2021.

\bibitem{Pandya2017}
Shishir Pandya, Joshua~D Wilbur, Bikram Bhatia, Anoop~R Damodaran, Christian
  Monachon, Arvind Dasgupta, William~P King, Chris Dames, and Lane~W Martin.
\newblock {Direct Measurement of Pyroelectric and Electrocaloric Effects in
  Thin Films}.
\newblock {\em Physical Review Applied}, 7(3):034025, mar 2017.

\bibitem{Glaum2012}
Julia Glaum, Yuri~A. Genenko, Hans Kungl, Ljubomira {Ana Schmitt}, and Torsten
  Granzow.
\newblock {De-aging of Fe-doped lead-zirconate-titanate ceramics by electric
  field cycling: 180°- vs. non-180° domain wall processes}.
\newblock {\em Journal of Applied Physics}, 112(3):034103, aug 2012.

\bibitem{Chen2021}
Chuan Chen, Yan Wang, Zong-Yue Li, Chun Liu, Wen Gong, Qing Tan, Bing Han,
  Fang-Zhou Yao, and Ke~Wang.
\newblock {Evolution of electromechanical properties in Fe-doped
  (Pb,Sr)(Zr,Ti)O$_3$ piezoceramics}.
\newblock {\em Journal of Advanced Ceramics}, 10(3):587--595, jun 2021.

\bibitem{Tong2014a}
Trong Tong, J. Karthik,R. V. K. Mangalam, Lane W. Martin, David G. Cahill,
\newblock {Reduction of the electrocaloric entropy change of ferroelectric PbZr$_{1-x}$Ti$_x$O$_3$ epitaxial layers due to an elastocaloric effect}.
\newblock {\em Physical Review B}, 90(9):094116, sep 2014.

\bibitem{Pandya2018b}
Shishir Pandya, Gabriel Velarde, Lei Zhang, and Lane~W. Martin.
\newblock {Pyroelectric and electrocaloric effects in ferroelectric
  silicon-doped hafnium oxide thin films}.
\newblock {\em Physical Review Materials}, 2(12):124405, dec 2018.

\bibitem{Lin2024}
Jiamin Lin, Fu~Lv, Zijian Hong, Bing Liu, Yongjun Wu, and Yuhui Huang.
\newblock {Ultrahigh Piezoelectric Response Obtained by Artificially Generating
  a Large Internal Bias Field in BiFeO$_3$ –BaTiO$_3$ Lead‐Free Ceramics}.
\newblock {\em Advanced Functional Materials}, 34(22):2313879, may 2024.

\end{thebibliography}

\end{document}